\documentstyle[prb,aps,floats,epsf]{revtex}
\begin{document}
\draft
\twocolumn[
\begin{center}
{\large\bf Flux lattice melting and depinning in the weakly frustrated 2D XY
model}\\
\vspace{3ex}
S{\o}ren A. Hattel and J. M. Wheatley$^*$\\
{\em Interdisciplinary Research Centre in Superconductivity,
University of Cambridge,} \\
{\em Madingley Road Cambridge, CB3 OHE, United Kingdom. }\\
( \today )
\end{center}
\widetext
\begin{abstract}
\leftskip 54.8pt
\rightskip 54.8pt

Monte Carlo simulations of the frustrated 2D XY model were carried out
at small commensurate values of the frustration $f$. For $f=1/30$ a single
transition was
observed at which phase coherence (finite helicity modulus) and vortex
lattice orientational order vanish together. For $f=1/56$ a new phase in
which phase coherence is absent but orientational order
persists was observed. Where comparison is possible, the results are in
detailed agreement with the behavior of
the lattice Coulomb gas model of vortices. It is argued that the helicity
modulus
of the frustrated 2D XY model vanishes for any finite temperature
in the limit of weak frustration $f$.

\end{abstract}
\leftskip 54.8pt
\rightskip 54.8pt

\narrowtext
]

Motivated by high temperature superconductors, there has been renewed interest
recently in the nature of phase
transitions, if any, in strongly type II 2D superconductors in applied magnetic
fields \cite{review}.
One approach to the 2D problem in the infinite $\kappa$ limit
is provided by the lowest
Landau level approximation to
Ginzburg-Landau theory, where a truncated basis consisting only
of states in the lowest
Landau level of linearized Ginzburg-Landau theory is retained \cite{lll}.
First order melting transitions have been observed in Monte
Carlo simulations using this
approach \cite{lllperiodic}, although no transition was observed
in simulations on a sphere \cite{lllsphere}.
An alternative discretization
procedure
places Ginzburg-Landau functional on a lattice, usually by
use of {\it phase only} models such as frustrated XY or Villain models.
Here fluctuations in the amplitude of the superconducting order parameter
are neglected, which is a reasonable approximation far
from the critical temperature. In the XY and Villain models
compactness (equivalence of states whose local superfluid phases differ by
multiples of $2 \pi$)
has been imposed in different ways.

The interaction potential of the Villain model has very special
features. The model shows
a separation of vortex and spin wave degrees of freedom; namely, after duality
transformation, the
partition sum factorizes into a product of spin wave (gaussian) fluctuations
and 2D lattice Coulomb gas components \cite{villain}. The lattice Coulomb gas
amounts to a 2D ``lattice London model" for the vortex state.
It was on the basis of the
continuum Coulomb gas model that Huberman and Doniach \cite{doniach} and
Fisher \cite{fisher} discussed
the vortex lattice melting transition in 2D superconductors.
Here Kosterlitz-Thouless dislocation mediated melting theory \cite{BKT,disloc}
leads to a melting temperature estimate which is
independent of vortex density. However,
in a series of papers Moore \cite{moore},
has implicitly questioned the validity of the London model, and
argued that the lower critical dimension for superconductivity is in fact
$d=4$.

The question naturally arises whether the phase only
discretization of Ginzburg-Landau theory provided by the
Villain model is fully equivalent to other discretization schemes.
If it were not, then the validity of conclusions based
on the Coulomb gas (London) model would be in doubt. For example,
in the alternative discretization
provided by the XY model,
the separation of vortex and spin-wave degrees
of freedom is not complete. Thus the London model is not strictly
valid even for the vortex
contribution to the thermodynamics.
Both the lattice London
and XY models have been extensively \cite{hsh} used in studies of vortex
lattice melting, and it is
essential to know if they indeed share the same continuum limit.

In a recent Letter \cite{teitelpreprint} Franz and Teitel reported very
detailed Monte Carlo simulations
of the lattice Coulomb gas (LCG) in two dimensions.
While the
discreteness of the lattice is expected to become unimportant in the
dilute limit,
these authors found that the ``continuum limit'' is subtle,
and only reached for
surprisingly small values of the density (denoted $f$).
In sufficiently dilute systems, a ``depinning" or ``floating''
phase transition occurs from a
low temperature phase with long range translational order (LRTO) to
a phase with
algebraic order, followed by a first order 2D melting transition to a
disordered phase.
For dense systems, the phase with algebraic order is absent, i.e.
depinning and
melting transitions coincide.
The existence of a phase with LRTO is
an artifact of pinning in the lattice model because there is a
finite energy cost for displacing a vortex in the ground state
configuration.
The appearance of a phase with algebraic order signals the onset
of the continuum limit.

The transition temperature $T_p$ at which
LRTO disappears vanishes linearly in $f$.
The 2D melting temperature,
where algebraic order disappears, is independent of
density as anticipated from 2D melting theory.
The linear dependence of $T_p$ on density in the
dilute Coulomb gas can be established analytically \cite{prb}.
The 2D {\it harmonic} Coulomb solid on a
lattice can be mapped to a dislocation
(vector plasma) problem, which has a continuous
Kosterlitz-Thouless-Halperin-Nelson \cite{BKT,disloc}
unbinding transition, provided that the core
energy for a dislocation is sufficiently large.
In the dilute limit the core energy can
be shown to grow as $\log(1/f)$ so that this
condition is always satisfied. In the low
temperature phase displacement fluctuations
of the Coulomb solid are screened which
permits LRTO even in 2D.

In this paper we report the results of a Monte Carlo study of the 2D XY
model in the limit of
weak commensurate frustration ($f=1/q$, $q={\mathrm {integer}}$)
with periodic boundary conditions. To avoid
biasing the vortex system towards hexagonal order, we choose
a square background grid. The 2D XY model is:
\begin{equation}
H = - \sum_{\langle i,j\rangle} \cos (\theta_i-\theta_j-A_{ij})
\label{hamiltonian}
\end{equation}
where $\theta_i$ is the phase of the superconducting order parameter defined
on an $L \times L$
square lattice of
points labeled by $i$.
The link
field $A_{ij}$ (line integral of the vector potential
between nearest neighbor lattice sites
$i$ and $j$) satisfies $\sum A_{ij} = 2 \pi \Phi / \Phi_0 \equiv 2 \pi f$.
The sum is taken counter clockwise around a plaquette.
The magnetic flux per plaquette $\Phi$ is uniform
which effectively imposes the infinite $\kappa$ limit.
$\Phi_0$ is the flux quantum, and $f$ is the frustration.
The local vorticity $\nu({\mathbf {R}})$
takes values $0, \pm 1$ and is defined via
$\sum {\hbox {mod}} (\theta_i - \theta_j - A_{ij}) = 2 \pi(f - \nu)$
where the sum is around the plaquette ${\mathbf {R}}$ \cite{korshunov}.
${\hbox {mod}}(x)$ adds integer multiples of $2 \pi$ to $x$ to
bring it into the range $(-\pi,\pi]$.
With periodic or anti-periodic
boundary conditions the total vorticity is constrained to
equal the total applied flux; i.e
the plaquette sum $\sum_{\mathbf R} \nu({\mathbf R}) = L^2 f$.

\begin{figure}
  \mbox{\leavevmode \epsfxsize=4.2cm \epsfbox{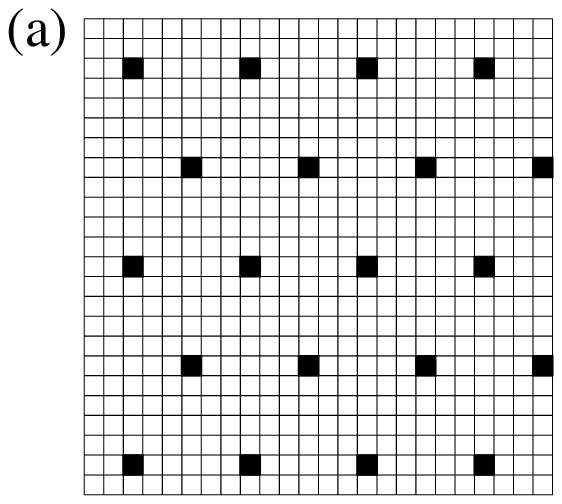} \hspace{1ex}
  \leavevmode \epsfxsize=4.2cm \epsfbox{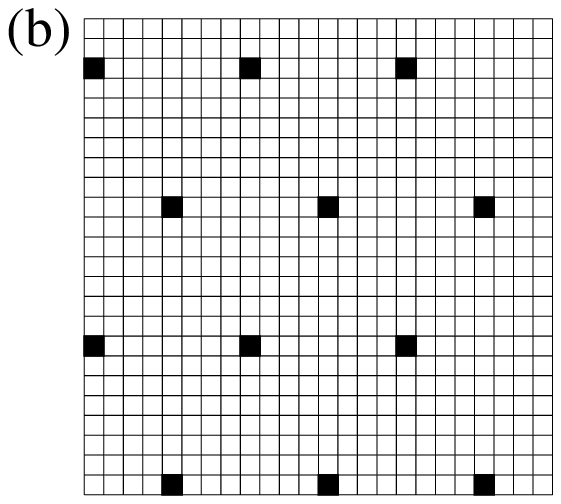} }
  \caption{The configuration of vortices in the ground states for
  (a) $f=1/30$, and
  (b) $f=1/56$. For convenience only system sizes $24\times 24$ are shown.}
  \label{groundstates}
\end{figure}
The ground state spin configurations of the XY
model in rational ($f=p/q$) magnetic fields change in
a highly non-trivial way with $p$ and
$q$ and have been the subject of a large literature \cite{ground}.
We have succeeded in finding ground states for
$f=1/30$ and $f=1/56$ appropriate to this study
by slow cooling of $q \times q$ unit cells, using periodic boundary
conditions for
$f=1/30$ and mixed periodic and anti-periodic conditions for $f=1/56$.
The vortex lattice is nearly triangular in these ground states
as shown in Fig.\ \ref{groundstates}. However, strictly speaking
these states have only two-fold symmetry.
Near hexagonal symmetry is associated with extra stability of these ground
states.

At finite temperatures we measure the helicity modulus and six-fold
orientational order parameter.
The helicity modulus, $\Upsilon$, is equivalent to the
superfluid density \cite{fbj}
and is a measure of long range phase coherence.
It is defined as the sensitivity of the free energy to a
twist in the boundary condition
along a particular direction via
\begin{equation}
\Upsilon = {1 \over L^2} {\partial^2 F \over \partial \delta^2} \mid_{\delta=0}
\label{eq:helicity}
\end{equation}
where $\delta$ is the twist angle \cite{helicity}.
Note that, for periodic boundary conditions,
$\Upsilon$ measures the free energy shift in the
presence of a flux loop of strength
$\delta$ about the interior
of the torus. The helicity modulus is therefore a
{\it gauge invariant} response.

In the continuum, a 2D solid has finite orientational order
below it's melting temperature.
The six-fold orientational order parameter is,
\begin{equation}
\varphi_6= {1 \over (f L^2)^2} \langle \sum_{k,l}
\exp[6 \imath (\phi_k - \phi_l)]\rangle
\label{eq:orient}
\end{equation}
where $\phi_k$ is the angle between a fixed direction in the
XY plane and the direction of the bond
between vortex $k$ and its nearest neighbor.
$\varphi_6 =1$ for a perfect triangular lattice.

We used a heat bath method for the simulations heating up slowly from
the ground states.
For $f=1/30$ we simulated systems with linear dimension
$L=60 and 90$, and for $f=1/56$
we simulated systems of size $L=56$ and $112$.
Before computing any averages the system
was carefully thermalized at each temperature discarding
$120,000-680,000$ Monte Carlo
sweeps (MCS) over the entire lattice. From these
thermalized states we computed four averages
using $20,000$ MCS, from which the final averages were calculated and the
error estimated from
the standard deviation. For the simulations we used several months
of CPU time on a HP7000
workstation. The task is more difficult than simulations of
the LCG, since there are $1/f$
times as many degrees of freedom involved in the updating
algorithm. Furthermore, the system sizes
accessible are restricted by the fact that the ground
states are only periodic on at best
$q \times q$ systems \cite{ground}.
For these reasons careful finite size scaling was
impractical.

A second important difference between XY and LCG models from
the point of view
of Monte Carlo simulation is the presence of energy barriers
to the motion of isolated vortices in the XY model.
To move a vortex through one lattice
constant requires finite motion of
spins in the vicinity of the vortex. This energy barrier has
been estimated to be
$E_p \sim 0.2$ \cite{pinning}, and plays an important role
in the physics of Josephson junction arrays.
The barrier means that
equilibration times become long at temperatures $T \ll E_p$.
Surprisingly, and for reasons not
completely clear to us, this problem was less severe than expected, and the
system equilibrates even at temperatures of order $0.1 E_p$,
as evidenced by our ability to
find ground states by cooling.
Further details of the Monte Carlo procedure and
additional results from these investigations will
be presented elsewhere \cite{toappear}.

\begin{figure}
   \epsfxsize=8cm \leavevmode \epsfbox{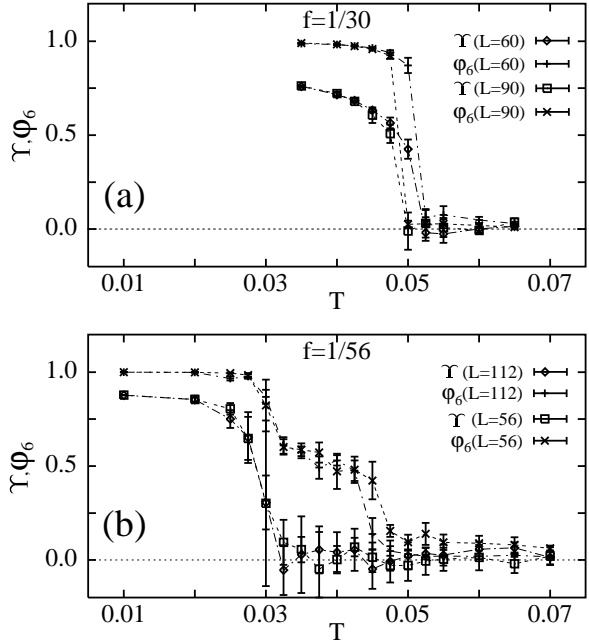}\\[3em]
   \caption{Orien\-tational order and helici\-ty modulus for
   (a) $f=1/30, L=60$ and $L=90$, (b) $f=1/56, L=56$ and
   $L=112$.
   Errorbars correspond to $\pm$ one standard deviation.
   Lines are guides to the eye.}
   \label{results}
\end{figure}
Fig.\ \ref{results} shows the computed curves for the
$\Upsilon$ and $\varphi_6$.
For $f=1/30$ orientational order and long
range phase coherence remain finite up to $T_p \approx 0.045$,
where they both drop rapidly to values around zero.
These data are consistent with a single $T_p$
at which both order parameters vanish.
For $f=1/56$ the helicity modulus vanishes at $T_p \approx 0.03$.
At this temperature
the orientational order also falls, but to
a clearly finite value. Orientational order
persists up to $T_m\approx 0.05$. The finite size effect is small.
Note that the melting temperature has not changed significantly
between $f=1/30$ and $f=1/56$.

As discussed above, to fully characterize
the three phases and
extract exponents for the two phase transitions is
a difficult task within the framework of XY model simulations.
However, it is possible to identify the phases with reasonable certainty
by analogy with the LCG, and by qualitative features of the structure function.
The results for $f=1/56$ clearly show the existence of two distinct phase
transitions. The first one at $T_p$ is
associated with the depinning of the vortex lattice from the underlying mesh.
The helicity modulus vanishes at the depinning transition, but
orientational order persists, consistent with the idea
that the vortices form a lattice
with at least algebraic translational order.

These results should be compared with
Franz and Teitel's \cite{teitelpreprint} LCG simulations
on a square lattice, who found that $T_p < T_m$ for $f \lesssim 1/30$
Their $f=1/60$ results are
equivalent, in our units,  to $T_p = 0.028$ and $T_m=0.047$.
Comparing Fig.\ \ref{results}
we see there is semi-quantitative agreement
between the behavior of the LCG model and the XY model. In
particular comparison with Fig. 1 of ref. [12] shows that
there is a striking similarity between the behavior of the
inverse dielectric function
of the LCG and the helicity modulus of XY model.
While the connection between $\epsilon^{-1}(T)$ and
$\Upsilon(T)$ can be rigorously established only for
the Villain model\cite{villain},
the close analogy between these two quantities is evident.

This interpretation is supported by the behavior of the
vortex lattice structure function,
\begin{equation}
  S({\mathbf q}) =  {1 \over L^2}
  \sum_{{\mathbf R}_i} \exp[\imath {\mathbf q}\cdot {\mathbf R}_i]
  \langle \nu({\mathbf R}_i) \nu(0) \rangle.
  \label{eq:nunu}
\end{equation}
\begin{figure}
  \begin{center}
  \mbox{\raisebox{2.5cm}{(a)} \vspace{-1em} \leavevmode \epsfxsize=3cm
  \epsfbox{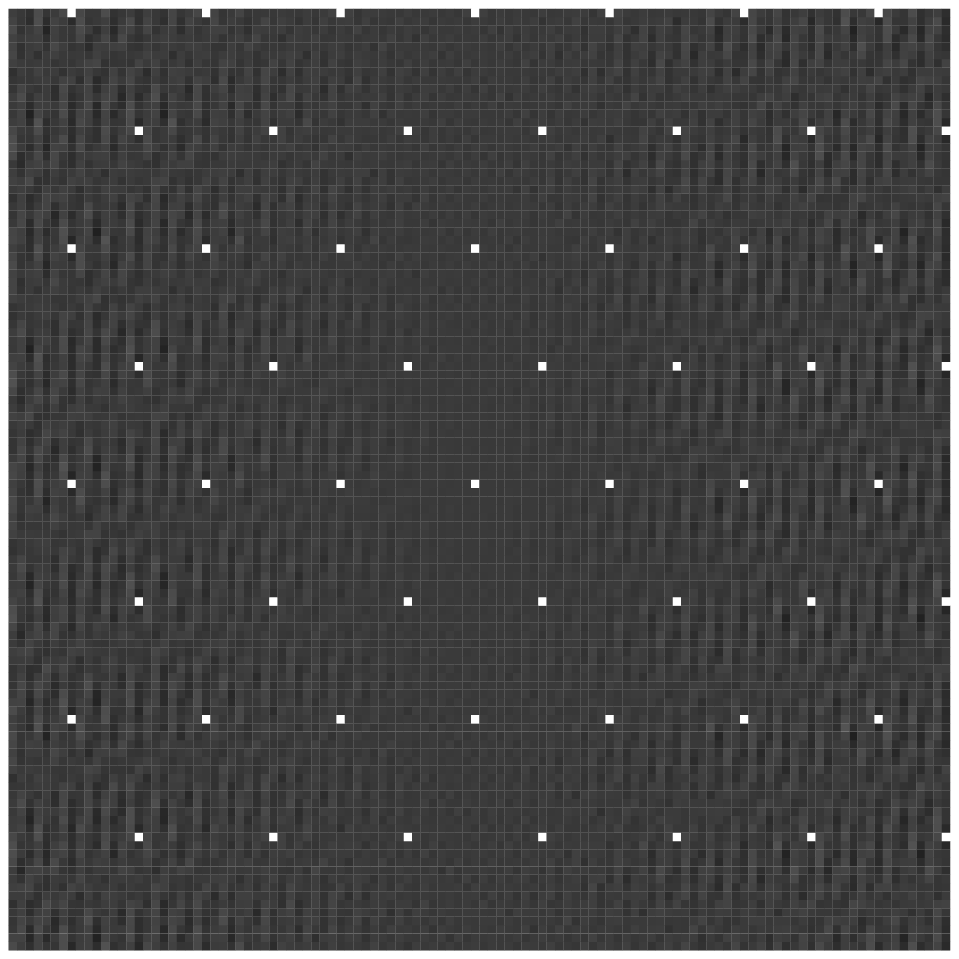}}
  \mbox{\raisebox{2.5cm}{(b)} \vspace{-1em} \leavevmode \epsfxsize=3cm
  \epsfbox{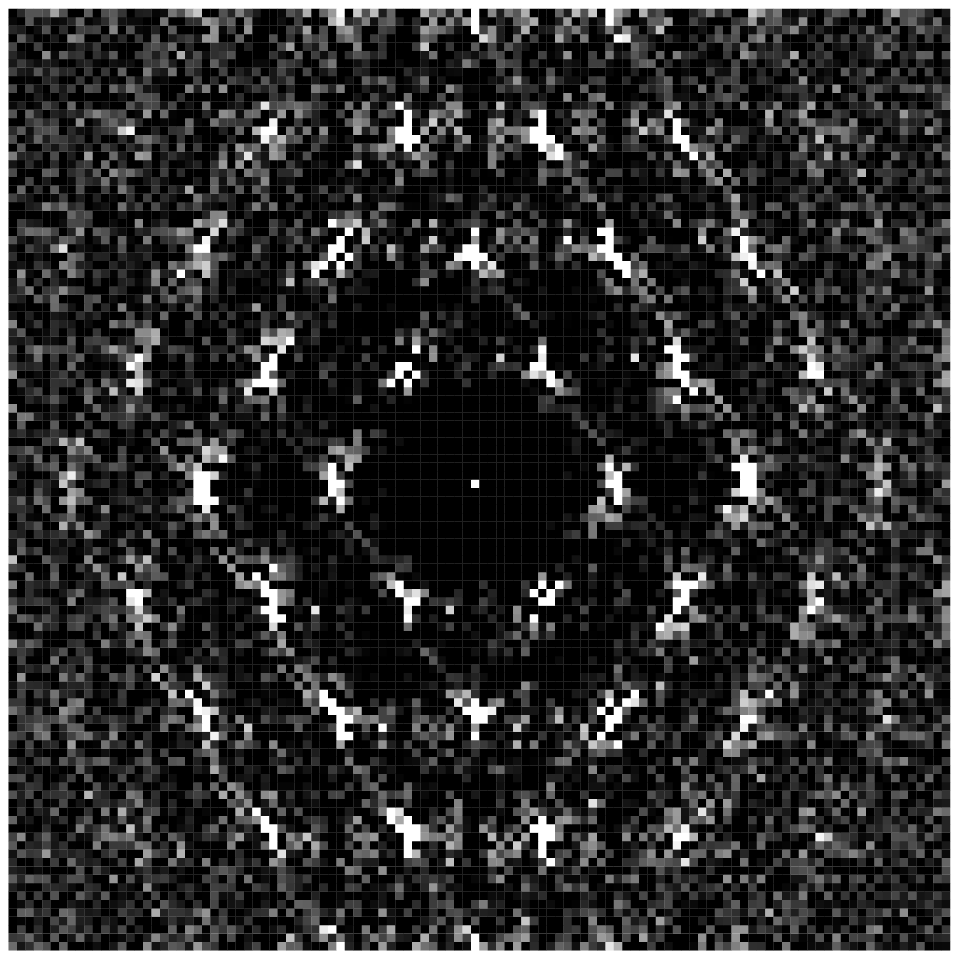}} \\
  \mbox{\raisebox{2.5cm}{(c)} \vspace{-1em} \leavevmode \epsfxsize=3cm
  \epsfbox{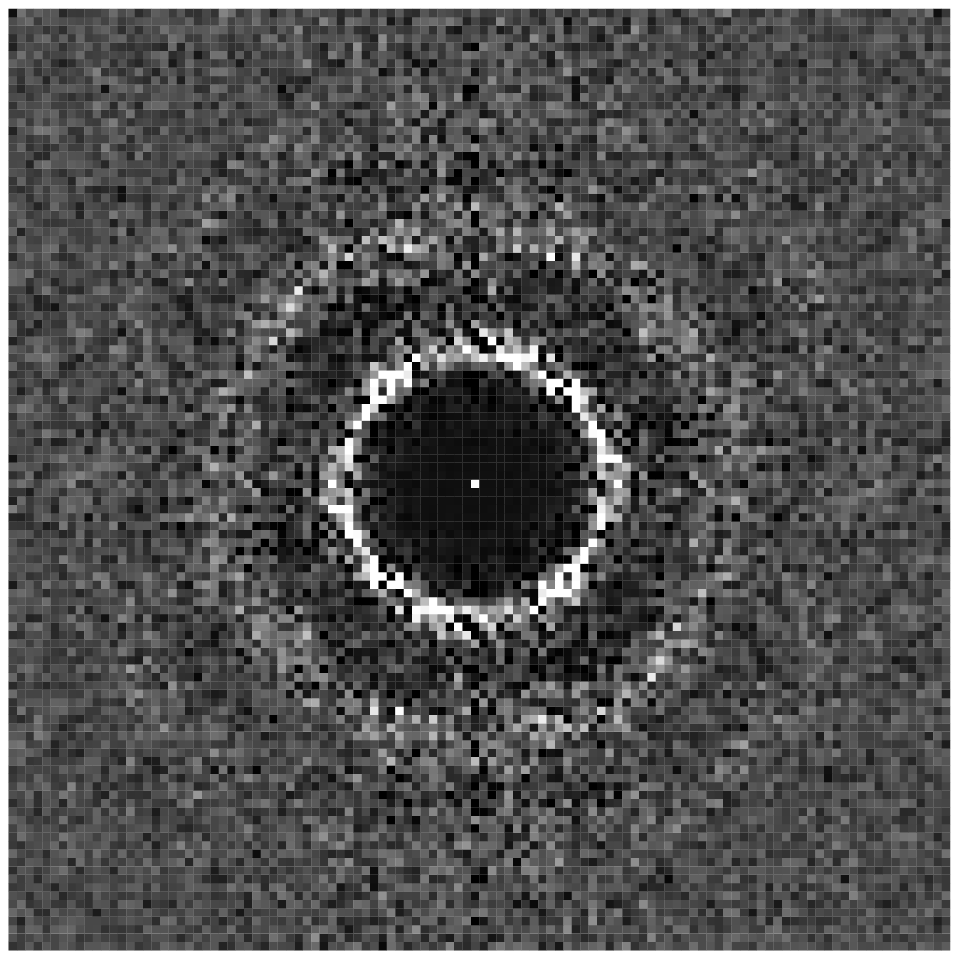}}
  \end{center}
   \caption{Vortex structure function in the square lattice Brillouin zone
   for $f=1/56$ and $L=112$ at temperatures (a) $T=0.02$,
      (b) $T=0.04$ and (c) $T=0.06$. }
   \label{spectra}
\end{figure}
Fig.\ \ref{spectra} shows
$S({\mathbf {q}})$ for $f=1/56$ for temperatures $T=0.02, 0.04 $
and $0.06$ corresponding to the three
phases of Fig.\ \ref{results}.b.
In the low temperature phase the Bragg peaks are
well defined out to high orders. In the intermediate temperature
phase, the
Bragg peak structure is still present, but the intensity of each
peak is much weaker, and
falls quickly with order. The near hexagonal symmetry remains clear,
compatible with algebraic order. In the high temperature phase the Bragg
peak structure is
absent and only the circular symmetry characteristic of a liquid is present.

Long range phase coherence disappears at the depinning transition,
leading directly to the conclusion that
the unpinned ideal vortex lattice with
algebraic order has no long range phase coherence
and is not superfluid \cite{ss,pins}.
Moreover, on the basis of the close analogy
between the LCG and frustrated XY models,
$T_p$ is expected to vanish linearly in the limit of small $f$
and thus the superfluid stiffness of the XY model vanishes
in this limit. This is remarkable because, as is well known, the
helicity modulus of the {\it unfrustrated} XY model is finite up
to the vortex unbinding
transition \cite{BKT} at $T_c=0.9$ \cite{xysim}. Here
the transition mechanism is the unbinding of vortex-antivortex
pairs and the helicity modulus shows a finite universal jump.
The origin of singular behavior in the small $f$ limit
is that a {\it depinned} Coulomb gas gives
rise to metallic screening of the
2D Coulomb vortex-anti-vortex interaction at long lengthscales, which
means that thermally excited
free vortices appear at any finite temperature, and the vortex
unbinding transition is absent.

It would be interesting to know the critical value of the
commensurate frustration,
such that for all $f<f_c$,
$T_p$ lies below $T_m$. However it appears unlikely to us that
$T_m=T_p$ for any $f<1/56$ because
$f=1/56$ is particularly strongly pinned, since it has near hexagonal symmetry.
It is also possible that depinning and melting transitions separate for
some values of
$f > 1/30$. While from our simulations we cannot rule out the presence
of an
hexatic phase for $T>T_m$, this was not
observed in the LCG \cite{teitelpreprint}.

In conclusion, the lattice Coulomb gas and
frustrated XY models show qualitatively very similar
melting and depinning phase
diagrams in low vortex density (weak commensurate frustration) limit.
In complete agreement with the arguments of Moore \cite{moore},
we find that thermal fluctuations of the vortex lattice destroy long range
phase coherence in the continuum limit of the
frustrated XY model
at any finite temperature. On the other hand, the model
shows a genuine thermodynamic phase transition, which can be
identified with the
disappearance of algebraic long
range order in the gauge invariant current pattern corresponding to the
vortex lattice structure.

We wish to thank M. Moore, A. Sudb{\o} and S. Teitel for helpful discussions.
SAH is supported by Carlsbergfondet.


\begin{references}
\bibitem{pins} We have also
        included isolated strong pinning centers in simulations for
	$f=1/56$ and found
        that $\Upsilon$ vanishes as before, i.e. pinning of a
	single vortex is not
        sufficient to pin a lattice with only algebraic order.
\end{references}
\end{document}